# Denoising with flexible histogram models on Minimum Description length principle

Vibhor Kumar
Laboratory of Computational Engineering
Helsinki University of Technology
P.O. Box 9203, FIN-02015 HUT, FINLAND
Email: vibhor@lce.hut.fi, Fax: +358-9-4514830

Jukka Heikkonen
Laboratory of Computational Engineering
Helsinki University of Technology
P.O. Box 9203, FIN-02015 HUT, FINLAND
Email: jukka.heikkonen@hut.fi

***Abstract*— Denoising has always been theoretically considered as removal of high frequency disturbances having Gaussian distribution. Here We relax this assumption and the method used here is completely different from traditional thresholding schemes. The data are converted to wavelet coefficients, a part of which represents the denoised signal and the remaining part the noise. The coefficients are distributed to bins in two types of histograms using the principles of Minimum Description Length. One histogram represents noise which can not be compressed easily and the other represents data which can be coded in small code length. The histograms made can have variable width for bins. The proposed denoising method based on variable bin width histograms and MDL principle is tested on simulated and real data and compared with other well known denoising methods.**

## I. Introduction

Traditionally it is assumed that there is a 'true' smooth signal to which noise, having almost always a Gaussian distribution, is added. Denoising is assumed to separate data $x^n = x_1, \ldots, x_n$, into a signal $\hat{x}^n = \hat{x}_1, \ldots, \hat{x}_n$ and 'noise' $e^n = e_1, \ldots, e_n$ thus:

$x_t = \hat{x}_t + e_t.$

In recent times orthonormal transforms have been used for denoising such as wavelets basis functions which combine powerful properties such as orthonormality, compact support, and varying degrees of smoothness and frequency with localisation in time and space.

Some proposed way of denoising using wavelet transform relies on retaining a number of the largest wavelet coefficients in absolute value, say coefficients $c_{(1)}, \ldots, c_{(k)}$ larger than a threshold, and setting the rest to zero, which after the inverse transform generate the estimate $\hat{x}^n$. Another traditional way has been so called soft-thresholding of coefficients. Such techniques with different thresholds are discussed in [1], [4], [5], [7], [10], [11], [12] and [19].

These thresholds method most often make assumption that the noise is the high frequency part in the data and that it has a Gaussian distribution or generalized Gaussian distribution with zero mean and some variance, which must be estimated from the data [1], [8], [2]. Some of the methods completely fail in absence of right estimation of noise variance, which itself is a typical task.

In contrast to these methods MDL based denoising [9], does not need any assumption of variance of noise. In [9] the data $x^n$ are modelled by a Gaussian distribution with mean $\hat{x}^n$, determined by a linear combination of the wavelet basis functions, the coefficients and the variance as parameters. It is a global Maximum Likelihood principle, global because it includes both the values of the parameters and their number as well as the structure where the parameters lie. Because of a constant that depends on the range and hence variance of the data is ignored, the criterion turns out to provide a superb denoising result up to a certain noise variance level but progressively worse when the variance increases.

A method of denoising without noise distribution assumption has been proposed in [13]. In [13] denoising is done by modelling both the wavelet coefficients representing the denoised signal and the rest representing the noise by equal bin-width histograms, for which the code lengths can be calculated. In [13] this method has been named as *MDL-histo* (MDL denoising with histograms). In broad terms MDL-histo can be described as: The wavelet coefficients are separated into a number of different resolution levels, each defining its own histogram. Instead of thresholds the selection of the retained coefficients on each resolution level is done by setting the coefficients in a subset of bins to zero, while leaving the coefficients intact in the remaining bins. The code length of the retained coefficients at each resolution level can be calculated. The code length for the noise is obtained by fitting a second histogram to the coefficients, obtained by setting to zero this time the retained coefficients in the set of all the coefficients that define the original data. The optimal set of bins is found by minimization of the sum of the two code lengths for the denoised signal and the noise. Although there are no thresholds in the proposed *MDL-histo* approach to denoising, the idea of 'soft thresholding' proposed in [13] has similar inspirations and similar effect as in [1], albeit in a manner which cannot be justified within the MDL theory. The problem with this MDL-histo method based on equal bin width is that there is always a chance of wrong grouping of coefficients in bins as coefficients representing noise and coefficients representing data may fall in one bin, so if that bin is dropped the achieved denoised signal may be too smooth. And sometimes a bin may not have any coefficients.

To solve the above problem with method proposed in [13] we extend it for variable bin width histograms. The theory of selecting the bins remain the same but we now use a better way to distribute coefficients in bins by utilizing variable bin widths.

This paper is organized such that we give the details of the denoising algorithm with a short description on histogram modelling and histogram based denoising MDL-histo, followed by a section on proposed extension of MDL-histo. In the result section, we compare the performance of extended MDl-histo with constant bin width MDL-histo and other denoising methods. In the final section we briefly draw conclusions.

## II. The MDL-histo algorithm

The code length $L$ for a data sequence $y^n = y_1, y_2, \dots, y_n$ of real valued data points $y_t$, quantized to a common precision $\delta$ and modelled by a histogram with $m$ equal width bins is given by

$$L(y^n|w, m, \delta) = \log \binom{n}{n_1, \dots, n_m} + \log \binom{n+m}{n} + n \log(w/\delta), \tag{1}$$

where all $y_t$ fall in the interval $[a, b]$, which is partitioned into $m$ equal width bins, the width given by $w = R/m$, where $R = b - a$, and $n_i$ data points fall in the $i$?th bin. The logarithms are to base 2, see [14] and [15].

The first term is the code length for encoding the $n$ bin indices corresponding to the bins of $y_1, y_2, \dots, y_n$. The second term is the code length for the integers $n_i$. To see this notice that the positive indices can be encoded as a binary string, which starts with $n_1$ zeros and a 1, followed by $n_2$ zeros and a 1, and so on. The string has $m$ 1's and its length is $n + m$. If we sort all such strings we can encode the index of any one with code length given by the logarithm of their number $\binom{n+m}{n}$. The third term gives the code length for the numbers $y_t$, quantized to the precision $\delta$. Indeed, each quantized number is one of the $w/\delta$ points in its bin. If we add the code length $L(m) = \log m + 2 \log \log m$ for the number $m$ we can minimize the conditional code length

$$\min_m \{L(y^n|w, m, \delta) + \log m + 2 \log \log m\}$$

conditioned on the two numbers $w$ and $\delta$.

We need to consider the code length of the modified data string $\hat{y}^n$, obtained by retaining the data points in a subset $S$ of the bins while the rest are set to zero; i.e., the points in the set of the remaining bins $\bar{S}$ are 0. Denote the indices of the bins of the data points $\hat{y}_t$ by $(0), (1), \dots, (|S|)$, where $|S|$ denotes the number of the bins in the set $S$. The first is the index of the bin that contains all the $n - k$ 0-points, where $k$ denotes the number of the retained points falling in the bins of $S$. The code length for the sequence of the bin indices is now

$$\binom{n}{n_{(1)}, \dots, n_{(|S|)}, n-k}, \tag{2}$$

where $n_{(j)}$ denotes the number of points falling in the bin having the index $(j)$. For instance, if the fifth bin is the first retained bin, then $(1) = 5$. In order to encode the sequence we need to know the bin indices of the points of $y^n$, not only that they belong to the first, second, etc bins of $S$. The code length for the string $\hat{y}^n$, given $w$, $m$, and $\delta$, is

$$L(\hat{y}^n|m, w, \delta) = \log \binom{n}{n_{(1)}, \dots, n_{(|S|)}, n-k} + \log \binom{n+|S|+1}{n} + k \log(w/\delta) + m. \tag{3}$$

The last term is the code length for $S$ as there are $2^m$ subsets $S$.

Using the above criterion denoising method using equal bin width histograms for different layers of wavelet, has been suggested in [13]. The criterion to minimised at $ith$ wavelet layer has been derived to be

$$\begin{aligned}
&\min_{S_i, M} \{ \sum_{j=1}^{i-1} \{ \log \binom{n_j}{n_{j,(1)}, \dots, n_{j,(m_j)}, n_j - k_i} \\
&+ \log \binom{n_j + m_j + 1}{n_j} \\
&+ \log \binom{n_i}{n_{i,(1)}, \dots, n_{i,(m_i)}, n_i - k_i} + \log \binom{n_i + m_i + 1}{n_i} \\
&+ \log \binom{n - \sum_{j=1}^{i-1} \hat{k}_j - k_i}{\nu_{i,1}, \dots, \nu_{i,M}} + \log \binom{n + M - \sum_{j=1}^{i-1} \hat{k}_j - k_i}{M} \\
&+ (n - \sum_{j=1}^{i-1} \hat{k}_j - k_i) \log(R_e/M) + (\sum_{j=1}^{i-1} \hat{k}_j + k_i) \log(R/m) \\
&+ \log(R_e) + \log \log(R_e) \log M + 2 \log \log M \},
\end{aligned} \tag{4}$$

where $m_j$ represents the number of selected bins from the $jth$ layer wavelet coefficients histogram and $\hat{k}_j$ is the number of selected coefficients from the $jth$ layer. It is clearly a two part coding here as model part consisting of different coded histogram has been coded separately and the residual part has been coded differently with M bin histograms with bins $\nu_{i,1}, \dots, \nu_{i,M}$. All the ranges of different layer histograms in the model part have been taken as one value equal to largest range among all the layer coefficients. The range $R_e$ of the residual part keep on varying.

Similar derivation can be made for variable bin width histogram model where the code for bin's end points can also be a part of the code length. The bin's beginning $a_i$ and end point $b_i$ can be optimised using the method described in [15] as the stochastic complexity of the retained bin would be completely different now. This would need large computational time to find the optimal bin limits with choosing of retained coefficients. Another way to achieve variable bin width histograms is to assume that the distribution of the wavelet coefficients in some functional term so that the bin boundaries can easily be calculated and the coder needs less codes for the boundaries. So if we ignore the small code length for the bin boundaries assuming functional form for the coefficients distribution the criterion in eq. 5, can be used with variable bin width histograms also. In order to implement this we used equal bin size histogram. Making equal bin width histogram with wavelet coefficients can be done by approximating the wavelet coefficient distribution with a function. By analysing wavelet coefficients of many types of data we concluded that most of the wavelet coefficients, specially representing the high frequency have laplacian distribution. Let $i\delta$ represent the quantisation of wavelet coefficient $c_i$, hence the approximately equal bin size or variable bin width histogram can be calculated using the assumed Laplacian distribution for coefficients $(i\delta)$

$$f(i\delta) = (1 - e^{-\lambda i\delta})e^{-\lambda i\delta} \tag{5}$$

keeping constraint $\sum f(i\delta) = 1$, where the deviation  can be estimated from the coefficients using the maximum likelihood approach to be

$$\lambda_{MLE} = (\log(1 + \delta/\overline{c}))/\delta, \tag{6}$$

where $\overline{c}$ is the mean of coefficients.

The optimisation to get minimum code length can be done over all the bins of the histograms of all the layers simultaneously. We call this optimisation as global optimisation. Unlike the work done in [13], the ranges used for different layers were their original ranges of their coefficients. After taking the corresponding ranges for different layers the criterion to be minimised by global optimisation for all the wavelet layers has been derived to be

$$\begin{aligned} L \min_{S_i,M} \sum_{j=1}^{L} \Big\{ & \log \binom{n_j}{n_{j,(1)}, \dots, n_{j,(m_j)}, n_j - k_j} \\ + & \log \binom{n_j + m_j + 1}{n_j} + \sum_{(i)=(1)}^{(m_j)} \log(w_{j,(i)}) \Big\} \\ + & \log \binom{n - \sum_{j=1}^{L} k_j}{\nu_{e,1} \dots, \nu_{e,M}} + \log \binom{n + M - \sum_{j=1}^{L} k_j}{M} \\ + & \sum_{l=1}^{M} \log(w_{e,l}) \}, \end{aligned} \tag{7}$$

Note that the bin width's $w_{j,(i)}$ for $jth$ layer $(i)$ bins are no longer equal . These bin widths are now calculated so that the number of coefficients lying in these bins are approximately same. Similarly the residual histogram bin widths $w_{e,l}$ are not all equal but their sizes are approximately the same.

Another form of code length that can also be used for this purpose is

$$\begin{aligned} \min_{S_i,M} \{ \sum_{j=1}^{L} \Big\{ & [\log \binom{k_j}{n_{j,(1)}, \dots, n_{j,(m_j)}} \\ + & \log \binom{k_j + m_j}{k_j} + \sum_{(i)=(1)}^{(m_j)} \log(w_{j,(i)}) \Big\} \\ + & \log \binom{n}{\nu_{e,1} \dots, \nu_{e,M}, n - \sum_{j=1}^{L} \hat{k}_j} + \log \binom{n + M + 1}{M} \\ + & \sum_{l=1}^{M} \log(w_{e,l}) \}, \end{aligned} \tag{8}$$

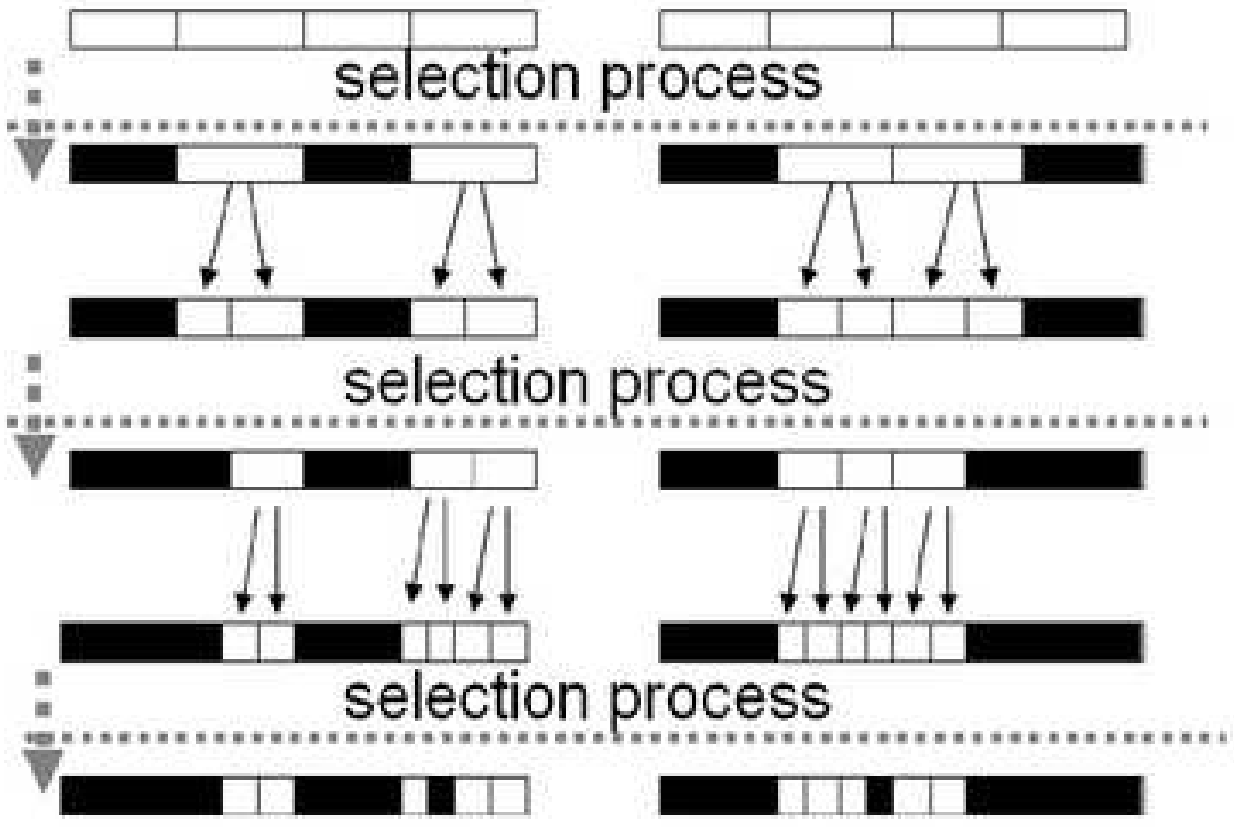

Fig. 1. The optimisation done with division of bins in to smaller size bins to get minimum code length

Notice here that the code length for location of non-selected coefficients in the model part has been replaced by code length for location of non-selected coefficients in the residual part.

The minimisation done earlier with equal bin width histograms tends to loose more coefficients than minimisation done with equal bin size coefficients and tend to blur the edges in the image denoising. This gives hint that there is possibility of further refinement for denoising, but use of higher number of bins has been limited by computational time. In order to do further refinement we implemented a method of successive picking of more coefficients after dividing the non-picked bins in to two bins, while keeping the coefficients from already picked bins. The implementation is represented in Fig. 1.

The simulation of successive picking method showed that by further refinement we can reduce the code length also, which satisfies the credibility of this method in terms of minimum code length principles. The ideal method to do global optimisation would have computational complexity is order of $\bigcirc(2^{mL})$, where $L$ is the total number of wavelet layers and $m$ is the number of bins in each layer histograms. The global optimisation can also be done using greedy approaches. The greedy approach that we suggest to use here for Gaussian noise is to sort all the bins according to coefficients magnitude and start selection process with bins with highest magnitude. Unlike thresholding method the higher magnitude bins should not be retained if their retention increases the code length. For other types of noises we found that starting selection process from low frequency wavelet layer histogram bins may lead to better results. But in both cases the overall selection process should be repeated for few iterations. Simulations have shown that these greedy approaches result almost similar results like ideal method.

## III. Experimental Results

Here we test our modifications to the MDL-histo method by applying it to simulated data with different types of noise, and the results are compared with the original MDL-histo and other denoising methods. We have applied Daubechies (db5), [17], wavelet basis functions in all the cases and the greedy approaches have been used to do the global optimisation while choosing the bins to get minimum code length.

As shown in Fig.(2a) time series signal has 4000 sample points consisting of two parts, a ramp, sinusoidal and square function, to which Gaussian and nongaussian noise are added. First, low variance Gaussian noise of zero mean and standard deviation 10 was added (Fig.2a) and a number of denoising techniques were tested, including BayesShrink [1], Fixedform thresholding method [6], Rissanen's linear-quadratic MDL-LQ denoising method [9], and previously proposed MDL-histo method with equal bin width histograms.

In addition to comparing the results through qualitative visual inspection we have calculated the Mean Absolute Error (MAE) measure for compared denoising techniques. Based on the results shown in Fig. 2a, Bayeshrink method seems to perform worst on one dimensional data. The MDL-histo with equal bin width histogram and the proposed MDL-histo with variable bin width histograms perform quiet similarly for Gaussian noise and the Fixedform method sometimes gives result equivalent to MDL-histo method with fixed bin width for low variance Guassian noise. When the noise distribution changes from Gaussian to nongaussian as shown in Fig(2b) Bayaeshrink and MDL-LQ did not give any good results so their results have not been shown. Fixedform method which gave comparative results with both MDL-histo methods for Gaussian noise does not tend to perform comparable to both MDL-histo methods, while MDL-histo method with variable bin width tend to produce better results than MDL-histo method with equal bin width. Of course the results shown here are preliminary, but some conclusion can still be made.

The MDL-histo methods with fixed bin width and variable bin width were tested on an cryo-electron microscope image of LDL particles, taken by a 200 KV electron microscope. The original figure shown in Fig. 3a shows the noise due to vitrified water around the sample. The distribution of noise is unknown but the variance of noise seems to be high. While there was no soft thresholding applied with both MDL-histo methods the Bayesshrink method was used with soft thresholding as the hard thresholding results were left out with lot of noise. Even after soft thresholding the image filtered by Bayesshrink method is left with a remarkable amount of noise, which could create problem in further classification of images in alignment of LDL particle images for 3D reconstruction [20]. The image filtered by MDL-histo method with fixed bin width and equal largest range for all the wavelet layers is blurry with a lot of edge information missing. The modification for using variable bin width histograms and variable ranges for different wavelet layers proposed here clearly shows advantage in Fig(4c), as the background noise is suppressed without loss of much information from region containing the LDL particles.

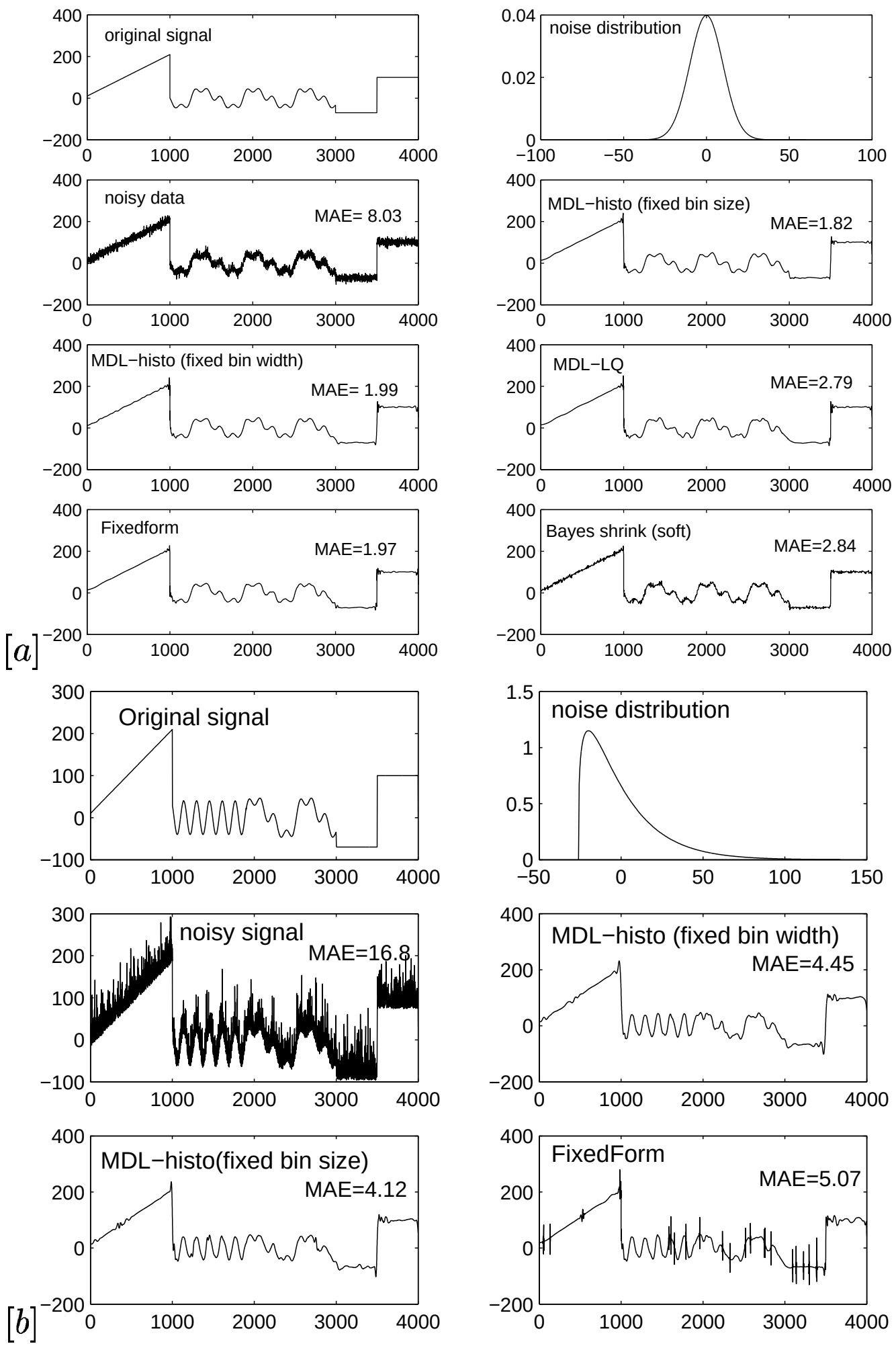

Fig. 2. Comparison of different denoising methods (a) The comparison of methods for Gaussian distributed zero mean noise (b) The comparison of methods on gamma distributed zero mean noise.

## IV. Conclusion

The histogram fitting after applying different orthonormal basis vectors to get two classes of histograms one for each orthonormal basis vector coefficients layer and another for non-selected coefficients can be used to separate the data in to two parts denoised and noise, which are different in terms of complexity. The main issue addressed in this paper is to use flexible histograms to model all types of noise with the principle of code length minimization to separate noise from data. Although the proposed MDL-histo technique is described in the wavelet domain, but histograms based two part coding can be used for denoising using other transforms which rely on energy compaction.

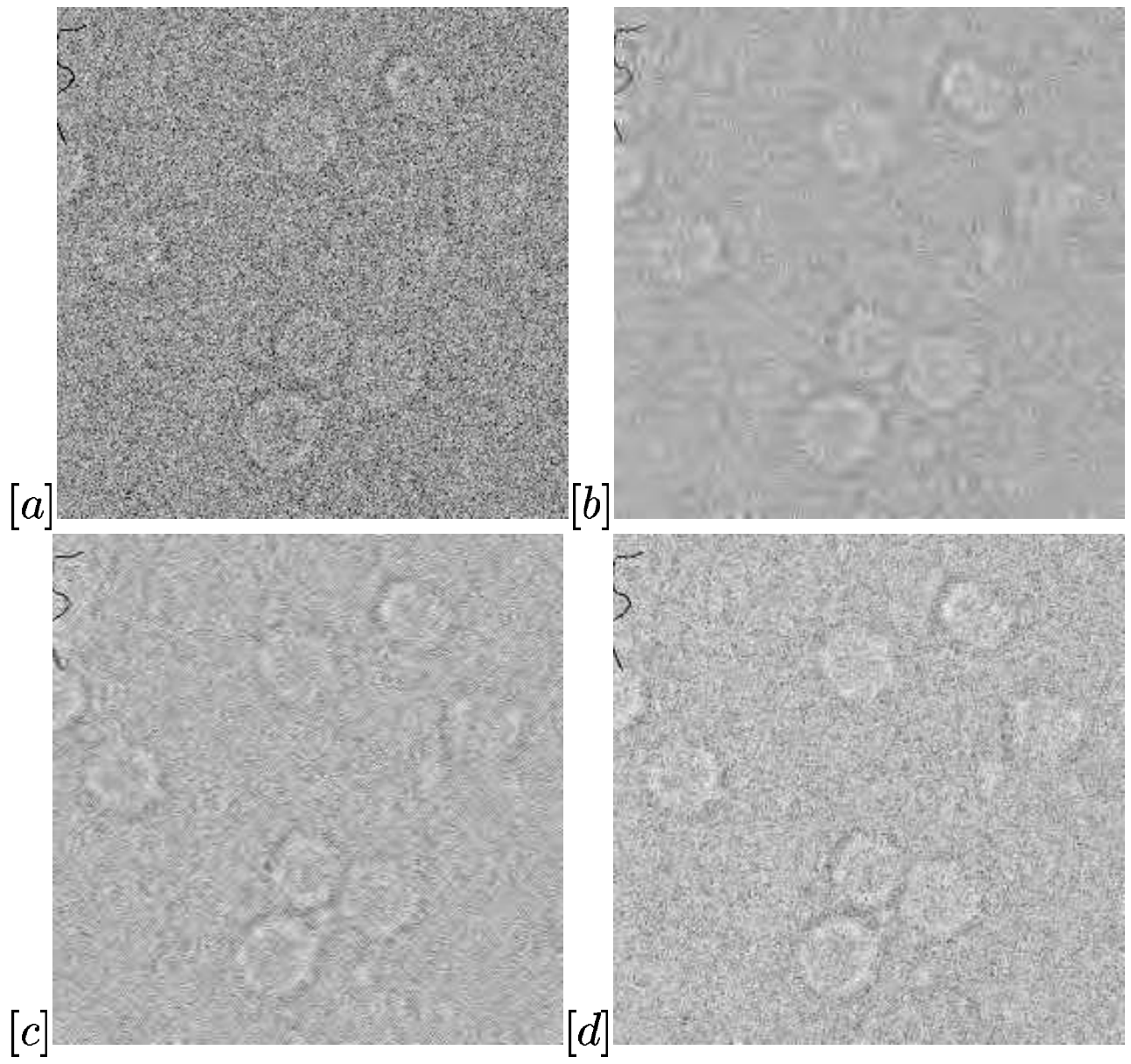

Fig. 3. Comparison of different denoising methods on cryo-EM images (a) Original noisy image rescaled to interval [0,1]. (b) MDL-histo with fixed bin width, (c) MDL-histo with variable bin width and approximately equal bin sizes (d) Bayeshrink method with soft thresholding.

## REFERENCES

[1] S. G. Chang, B. Yu and M. Vetterli, ”Adaptive Wavelet Thresholding for Image Denoising and Compression,” *IEEE Trans. on Image Processing*, vol.9, pp. 1532-1546, September 2000.

[2] E. Simoncelli and E. Adelson, ”Noise removal via Bayesian wavelet coring,” in *Proc. IEEE Int. Conf. Image Processing*, vol. 1, pp. 379-382, Sept. 1996.

[3] H. Krim and I. C. Schick, ”Minimax Description Length for Signal denoising and Optimized Representation,” *IEEE Trans.on Information Theory*, vol. 45, no. 3, pp. 898-908, April 1999.

[4] D. L. Donoho and I. M. Johnstone, ”Ideal Spatial adaptation by wavelet Shrinkage,” *Biometrika*, vol. 81, pp. 425-455, 1994.

[5] D. L. Donoho and I. M. Johnstone,”Adapting of Unknown Smoothness via Wavelet Shrinkage,” *Journal of American Statistical Assoc.*, vol. 90, no.432, pp. 1200-1224, December 1995

[6] D.L.Donoho, ”Progress in wavelet analysis and WVD: a ten minute tour,” in *Progress in wavelet analysis and applications*, Y. Meyer, S. Roques, Ed. FrontiA res , 1993, pp. 109-128.

[7] Abarmovich, T. Sapatinas, and B. W. Silverman, ”Wavelet thresholding via a Bayesian approach,” *J. R. Statist. Soc.*, ser. B, vol.60, pp.725-749,1998.

[8] M. Antonini, M. Barlaud, P. Mathieu, and I. Daubechies, ”Image coding using wavelet transform,” *IEEE Trans. Image Processing*, vol. 1, no. 2, pp. 205-220, 1992.

[9] J. Rissanen, ”MDL Denoising,” *IEEE Trans. on Information Theory*, vol. 46, pp. 2537-2543, 2000.

[10] G. Nason, ”Choice of the threshold parameter in wavelet function estimation,” in *Wavelets in Statistics*, A. Antoniades and G. Pooenheim, Ed. Berlin, Germany: Springer-verlag, 1995.

[11] Y. Wang, ”Function estimation via wavelet shrinkage for long-memory data,” *Ann. Statist.*, vol. 24, pp.466-484,1996.

[12] H. Chipman and E. Kolaczyk, and R. McCulloch, ”Adaptive Bayesian wavelet shrinkage,” *J. American Statistical Association*, vol. 92, no. 440, pp.1413-1421, 1997.

[13] V. Kumar, J. Heikkonen, J. Rissanen and K. Kaski, ”Minimum Description Length denoising with histogram models,” *IEEE Trans. on Signal Processing in Press*,

[14] P. Hall and E: J. Hannan, ”On Stochastic Complexity and Nonparametric Density Estimation,” *Biometrika*, vol. 75, pp. 705-714, December 1988.

[15] J. Rissanen, T. P. Speed and B. Yu, ”Density estimation by stochastic Complexity,” *IEEE Trans. on Information Theory*, vol. 38, No. 2, pp. 315-323, march 1992.

[16] T.Roos, P.Myllymäki, H.Tirri, ”On the Behavior of MDL Denoising”, in *Proceedings of the 10th International Workshop on Artificial Intelligence and Statistics (AISTATS)*, Barbados, January 2005 (to appear).

[17] I. Daubechies, ”Ten Lectures on wavelets,” in *CBMS-NSF Regional Conference Series in Applied Mathematices*, Philadelphia, PA:SIAM, 1992.

[18] Lim, Jae S. ”Two-Dimensional Signal and Image Processing,” Englewood Cliffs, NJ: Prentice Hall, 1990. pp. 536-540.

[19] D. L. Donoho, ” De-noising by soft-thresholding,” *IEEE Trans. Inform. Theory*, vol. 41, pp.613-627, May 1995

[20] M van Heel ,B Gowen,R Matadeen ,EV Orlova,R Finn,T Pape,D Cohen,H Stark,R Schmidt, M Schatz,A Patwardhan, ” Single-particle electron cryo-microscopy: towards atomic resolution,” *Q Rev Biophys*, vol 33(4) pp.307-69, Nov 2000.